\documentclass[conference]{IEEEtran}
\IEEEoverridecommandlockouts
\usepackage{cite}
\usepackage{amsmath,amssymb,amsfonts}
\usepackage{algorithmic}
\usepackage{graphicx}
\usepackage{textcomp}
\usepackage{xcolor}
\def\BibTeX{{\rm B\kern-.05em{\sc i\kern-.025em b}\kern-.08em
    T\kern-.1667em\lower.7ex\hbox{E}\kern-.125emX}}
\begin{document}

\title{UAV Access Point Placement for Connectivity to a User with Unknown Location Using Deep RL\\
\thanks{This work was supported in part by the CONIX Research Center, one of six centers in JUMP, a Semiconductor Research Corporation (SRC) program sponsored by DARPA.}
}

\author{\IEEEauthorblockN{Enes Krijestorac, Samer Hanna, Danijela Cabric}
\IEEEauthorblockA{\textit{Electrical and Computer Engineering Department,} \\ \textit{University of California, Los Angeles} \\ 
%Los Angeles, USA\\
		enesk@ucla.edu,
		samerhanna@ucla.edu, danijela@ee.ucla.edu 
}
}

\maketitle

\begin{abstract}
In recent years, unmanned aerial vehicles (UAVs) have been considered for telecommunications purposes as relays, caches, or IoT data collectors. 
In addition to being easy to deploy, their maneuverability allows them to adjust their location to optimize the capacity of the link to the user equipment on the ground or of the link to the basestation.
The majority of the previous work that analyzes the optimal placement of such a UAV makes at least one of two assumptions: the channel can be predicted using a simple model or the locations of the users on the ground are known.
In this paper, we use deep reinforcement learning (deep RL) to optimally place a UAV serving a ground user in an urban environment, without the previous knowledge of the channel or user location. 
Our algorithm relies on signal-to-interference-plus-noise ratio (SINR) measurements and a 3D map of the topology to account for blockage and scatterers. Furthermore, it is designed to operate in any urban environment.
Results in conditions simulated by a ray tracing software show that with the constraint on the maximum number of iterations our algorithm  has a 90\% success rate in converging to a target SINR.
\end{abstract}

\begin{IEEEkeywords}
UAV, relay, IoT, reinforcement learning
\end{IEEEkeywords}

\section{Introduction}
Due to their high mobility and low cost, unmanned aerial vehicles (UAVs) have found their way to many applications in recent  years, including  package  delivery, law  enforcement, search and rescue, etc. Following this trend, UAVs are getting an increased attention in the telecommunications sector.
Deploying UAVs as aerial basestations has recently emerged as an idea to respond to high localized traffic demands in the next-generation cellular networks \cite{li2017uav, wu2018uav, mozaffari2016efficient}. Using UAVs in such way provides the opportunity to exploit their agility of motion to improve  the  air-to-ground  link capacity  by optimal air placement. 
%Furthermore, due to the fact that they can fly at high altitudes UAVs can achieve a line-of-sight (LOS) link to the core basestation more easily than the ground users. 
UAVs can also be utilized for data harvesting in IoT or as data caches and in these applications it is also important to maximize the air-to-ground capacity by optimal placement.   

In this paper, we are interested in optimizing the capacity of the channel between the UAV and a ground user in an urban environment. 
This is a challenging problem considering that the environment between the UAV and the ground equipment can be abundant in scatterers and therefore hard to account for analytically and numerically. 
Nevertheless, the said problem has been addressed in literature before, under different assumptions. 
In most cases, however, the solutions are based on the assumptions that the ground user locations are known and that the wireless channel can be predicted with a simple model. 

The problem of a UAV relay placement has been considered mostly for line-of-sight (LOS) channels. In \cite{wang2018unmanned, jin_joint_2012, zeng_throughput_2016} transmit power and placement of a UAV relay are jointly optimized. 
However, the authors' solutions apply only to a LOS propagation channel, which makes this approach less applicable in the environments with scattering and obstacles, such as cities. 
Furthermore, all of \cite{wang2018unmanned, jin_joint_2012, zeng_throughput_2016}  assume that the locations of the users and channel propagation characteristics are known. 
In \cite{muralidharan2017path}, a method for optimizing the location of a ground unmanned vehicle is proposed. The approach relies on user location and assumes a fading channel. The algorithm predicts the channel quality across the entire map from a small number of measurements and then using stochastic dynamic programming an unmanned vehicle is optimally routed. While this paper considers ground vehicles, it is relevant to our work since their approach can apply to aerial vehicles flying at a fixed altitude. 

In addition to statistical models, some of the previous approaches have also utilized topology maps. The work in \cite{ladosz2016optimal} considers a UAV swarm that relays communication between users on the ground in an urban environment. 
The approach relies on known user locations and topology maps to perform swarm particle optimization of placement.
Works \cite{chen2017optimal,esrafilian2018learning,esrafilian2018uav} utilize 3D topology maps to help UAV placement optimization. 
Additionally, \cite{chen2017optimal} uses a statistical channel model and the user location to perform optimization. While \cite{esrafilian2018learning} does rely on user locations, it does not need to know the channel model parameters as these are learned. 
The algorithm in \cite{esrafilian2018uav} simultaneously learns user locations and parameters, but as a result, the approach has an extensive learning phase. 

Seeking to develop an exploration algorithm that performs learning and placement optimization simultaneously we turned to reinforcement learning. Reinforcement learning has been previously applied to similar problems. \cite{liu2018deployment} uses table-based Q-learning for optimal placement of aerial basestations with the knowledge of user locations. 
However, since the inputs to this algorithm are only user and UAV locations, it cannot perform outside of the environment it has been trained on. 
Similarly, \cite{chowdhury2019rss} uses received signal strength at the UAV and the UAV location to track indoor users with a shallow Q-learning algorithm. However, the paper only considers two indoor scenarios and the training and testing dataset are the same. Therefore, it is not clear whether the algorithm could perform in a new environment.  

We address the problem of optimal UAV placement assuming that the user location is not known. 
Algorithms that rely on statistical models of the channel may fail to generalize to all environments since the local topology can significantly differ from statistical predictions.
To that end, we use deep reinforcement learning to obtain a model-free algorithm for UAV positioning. 
The proposed algorithm relies on the knowledge of local topology and does not require the knowledge of the user location. Deep reinforcement learning allows us to take a high-dimensional input that is the topology map and use it alongside SINR measurements collected on the trajectory of the UAV to predict the optimal direction of motion. We test our performance in a realistic environment that emulates the wireless channel using a ray-tracing software. Furthermore, the training and the testing dataset are different. 

The rest of this paper is organized as follows. In Section \ref{sec:problem-casting} we introduce the relevant reinforcement learning background, define our problem and describe how reinforcement learning can be used to tackle it. In Section \ref{sec:dataset}, we describe the simulation environment. Sections \ref{sec:results} and \ref{sec:conclusion} are dedicated to results and conclusions, respectively.

\section{UAV Placement using Q-learning}
\label{sec:problem-casting}
In this section, we first introduce the necessary background in reinforcement learning in part \ref{sec:background}. Next, in subsection \ref{sec:fomulation}, we formulate our problem as a partially observable Markov decision process (PO-MDP) and in subsection \ref{sec:implementation} we discuss how we apply Q-learning to this PO-MDP.

\subsection{Reinforcement learning background}
\label{sec:background}
\newcommand{\mStep}{d}
\newcommand{\mAgent}{\mathcal{A}}
\newcommand{\mEnvironment}{\mathcal{E}}
%\begin{figure}[t!]
%	\centering
%	\includegraphics[width=0.45\textwidth]{figures/city.jpg}
%	\caption{The environment that was simulated to train the model. The size of the shown area is 960x590 m.}
%	\label{fig:training-ground}
%\end{figure}
Reinforcement learning (RL) is the branch of machine learning
that is concerned with making sequences of decisions. It is mainly concerned with problems that can be casted as a Markov decision process (MDP). 
In an MDP, an agent $\mAgent$ is situated in an environment $\mathcal{E}$. 
At each timestep $t$, the agent is in a state $s_t$ and takes an action $a_t$, it receives a reward $r_t$ while moving into the state $s_{t+1}$. 
In a partially observable MDP (PO-MDP), the full knowledge of the state of the environment is not known to the agent and in that case it will only have access to an observation of the state. This  observation then replaces the function of the state in the reinforcement learning algorithms.
%While the observation is sometimes denoted differently, in the remainder of this paper we will use the variable $s_t$ to represent the observation of the state.

The objective function in reinforcement learning is often the expectation of the discounted reward, $\mathbb{E}\sum_{t=0}^\infty \gamma^t r_t$, where $\gamma$ is the discount factor. 
Reinforcement learning methods can broadly be classified into two categories: policy learning and Q-learning. 
In policy learning methods, the goal is to learn the optimal policy function that defines $\pi(a|s)$, which is the probability of taking the action $a$ in the state $s$. The policy is often deterministic and in that case $\pi(a|s)$ defines a single action in each state. 
In Q-learning methods, the goal is to learn the Q-value function 
\begin{equation*}
Q(s, a) = \mathbb{E}[\sum_{t=0}^\infty \gamma^t r_t | s_0 = s, a_0 = a],    
\end{equation*}
that defines the expected reward in a state $s$, after taking the action $a$.
If the Q-value function is known, the optimal action in a state $s$ is then ${\text{argmax}_a}~Q(s,a)$. 
Deep Q-learning is an extension to the Q-learning paradigm that uses deep learning models, such as convolutional neural networks and recurrent neural networks to approximate $Q(s, a)$.
Furthermore, Q-learning is a model-free reinforcement learning technique, meaning that it does not rely on known dynamics of the system.

One of the first deep Q-learning algorithms was proposed by Deepmind and was successfully demonstrated on Atari video games \cite{mnih2013playing}. The algorithm was named the deep Q-network (DQN) algorithm. 
In it, the Q-value function is parametrized by a neural network $Q_\theta$, with parameters $\theta$. 
An estimate of the true Q-value at time $t$, $Q_t$, can be obtained by using a single sample estimate of the Bellman backup operator
\begin{equation}
    \widehat{\mathcal{T}Q_t} = r_t + \underset{a_{t+1}}{\text{max}}~\gamma Q_\theta (s_{t+1},a_{t+1})
    \label{eq:bellman}
\end{equation}
This is called a single sample estimate because only the reward $r_t$ at the current time instant $t$ is used to approximate the infinite horizon Q-value function. 
%It has been shown that by repeatedly applying the backup operator the estimate will converge to the true value, with some noise \cite{jaakkola1994convergence}. 

In order to approximate $Q$ by $Q_\theta$ the following minimization is done over sample data,
\begin{equation}
    \underset{\theta}{\text{minimize}} \sum_t \left|\left| \widehat{\mathcal{T}Q_t} - Q_\theta(s_t, a_t)\right|\right|^2
    \label{eq:learn}
\end{equation}
In the DQN algorithm, the training and the interaction of the agent with the environment happen in parallel. 
As the agent gathers experience, samples of that experience are stored and the minimization in the Equation \ref{eq:learn} is done periodically, every $\tau_L$ steps, by randomly sampling a batch of $B_L$ recorded samples and applying gradient descent. 
This is referred to as experience replay. 
Samples are arrays of data $(s_t, a_t, r_t, s_{t+1})$ and these are stored in the replay buffer. 

The agent interacts with the environment following the $\epsilon$-greedy policy, where at any time $t$ the agent either takes a random action at probability $\epsilon$ or the Q-value optimal action $\text{argmax}_a~Q_\theta(s,a)$ at probability $(1-\epsilon)$. Over the course of the training, the value of epsilon decreases from 1 to 0. 
In the implementation of DQN, there is an additional $Q_\theta$, called the target Q-network. 
The target Q-network is used in the Bellman backup operator but it is not optimized over. Instead, it is periodically copied from the main Q-network. The target Q-network is included to improve the stability during training.

The original vanilla DQN algorithm has been improved upon over the years.
The two expansions that we will use are double Q-learning \cite{van2016deep} and dueling networks \cite{wang2015dueling}.
For the sake of brevity we omit the details of these algorithms and the reader is referred to the cited works for more information. 
%In these, the Q-value is parametrized as $Q(s,a) = V(s) + A(s,a)$, where $V(s) = \mathbb{E}[\sum_{t=0}^\infty \gamma^t r_t | s_0 = s]$ is the state-value function and $A(s,a) = Q(s,a) - V(s)$ is the advantage value function.  

\subsection{UAV placement problem as a PO-MDP}
\label{sec:fomulation}
We now formulate the UAV placement problem as PO-MDP. We consider the scenario of a UAV located in an urban environment communicating to a radio device on the ground. 
The area topology is such that LOS connection to the ground user is not always possible, and the communication will often occur over non-line-of-sight paths. 
At time zero, the UAV and the ground device are located at random positions and the goal of the UAV is to adjust its position so as to increase the SINR at the UAV.
We assume that the UAV receives some signal power from the user on the ground to begin with. The UAV moves until an SINR threshold is reached or until maximum time for optimization expires. 

In the following, we describe the mechanics according to which the UAV moves around. 
Since exploring the entire 3D space is a complex task, we restrict the motion of the UAV to the horizontal plane and assume that its altitude is kept constant. 
We do this as a relaxation but it is worth pointing out that the optimal position for a UAV will often be at the lowest allowed altitude since this brings it closest to the ground user. 
The UAV can only move around buildings or fly above buildings that are below its altitude and it makes adjustments in its position at discrete time steps. 
We restrict the directions of the motion of the UAV to the four orthogonal horizontal directions and the motion step size $d_S$ is fixed. 
We impose this constraint because Q-learning lends itself better to tasks with a discrete set of actions.
With these restrictions on the motion, the UAV effectively moves in a uniform plane grid space that spans the environment.

We assume that the UAV location is known.
Furthermore, the UAV has access to a 3D map of the environment that maps all the buildings, which can be drawn from a database. 3D maps of major urban areas are generally available and easy to acquire. We also make the assumptions that the channel is slow fading and that the user location does not change significantly over the course of optimization. Furthermore, we assume that there is a sufficient backhaul capacity between the UAV and the core basestation, so we only focus on optimizing the capacity between the UAV and the ground device.

\subsubsection{Observation space}
\begin{figure}[t]
	\centering
	\includegraphics[width=0.5\textwidth]{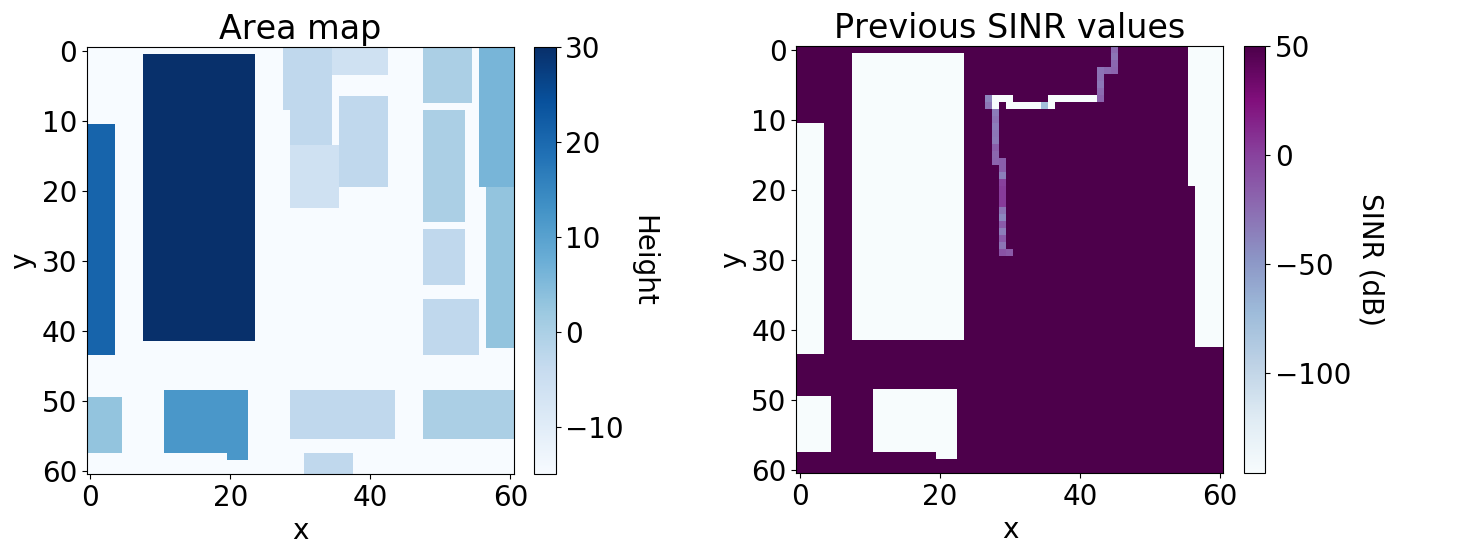}
	\caption{Example observation for an agent moving according to a random exploration policy. Note that the SINR values of areas that are obstructed and therefore unreachable are set to a very low value (-150 dB), while the areas that have not yet been visited are set to a very high value outside of the possible range of SINR values (50 dB).}
	\label{fig:observations}
\end{figure}

Since the full state of the system in which the UAV operates is not available, the agent in our algorithm relies on two types of observations of the environment to drive its decision making. The first type of observation is the 3D map of the local area.  The local area in our case is a square area of side $l_O$ centered at the current location of the UAV. The 3D map information is compressed into a 2D array representation, where each entry represents a grid point in the local area and the value of each entry corresponds to the height of the terrain at that point relative to the UAV altitude. We use heights relative to the UAV altitude to make the algorithm adjustable to different starting UAV flying heights.
%For coherence, we maintain the spacing of these grid points equal to the spacing of SINR measurements generated by channel simulation software. 

The second type of observation that the agent uses are the SINR measurements at previously visited locations in the local area. 
These are also stored in a 2D array with grid point locations matching the locations of the grid points in the topology observation. 
The value of each entry is the measured SINR value. To complete the array, we populate the entries with unknown SINR values with a high value $P_H$ outside of the regular range. Furthermore, the points that are blocked by buildings and cannot be visited are populated with low values $P_L$ outside of the span of possible SINR values. With successful training, the algorithm will learn the significance of $P_H$ and $P_L$. Example observations are shown in Figure \ref{fig:observations}.     

\subsubsection{Action space}
Since the action is the motion of the UAV at each time step, it can take on the values
(0, $d_S$), (0, -$d_S$), ($d_S$, 0), (-$d_S$,0), where each vector represents the displacement in the x- and y-coordinates.
\subsubsection{Reward}
The UAV receives a reward equal to the difference between the SINR in the next state and the SINR in the current state. 
This incentivizes the agent to move towards higher SINR.
Furthermore, we assign a constant exploration reward $c_E$ which the agent receives for visiting a new location. 
We empirically established that an exploration reward incentivizes the agent to explore further away from its starting location, which results in better performance. The reward is mathematically expressed as 
\begin{equation*}
r_{t}=\text{SINR}_{t+1}-\text{SINR}_{t}+c_E\delta_{t}^{E},
\end{equation*}
where $\delta_{t}^{E}$ is an indicator function activated when the UAV visits a new location.

\subsection{Deep Q-learning implementation}
\label{sec:implementation}
\begin{figure}[t]
	\centering
	\includegraphics[width=0.46\textwidth]{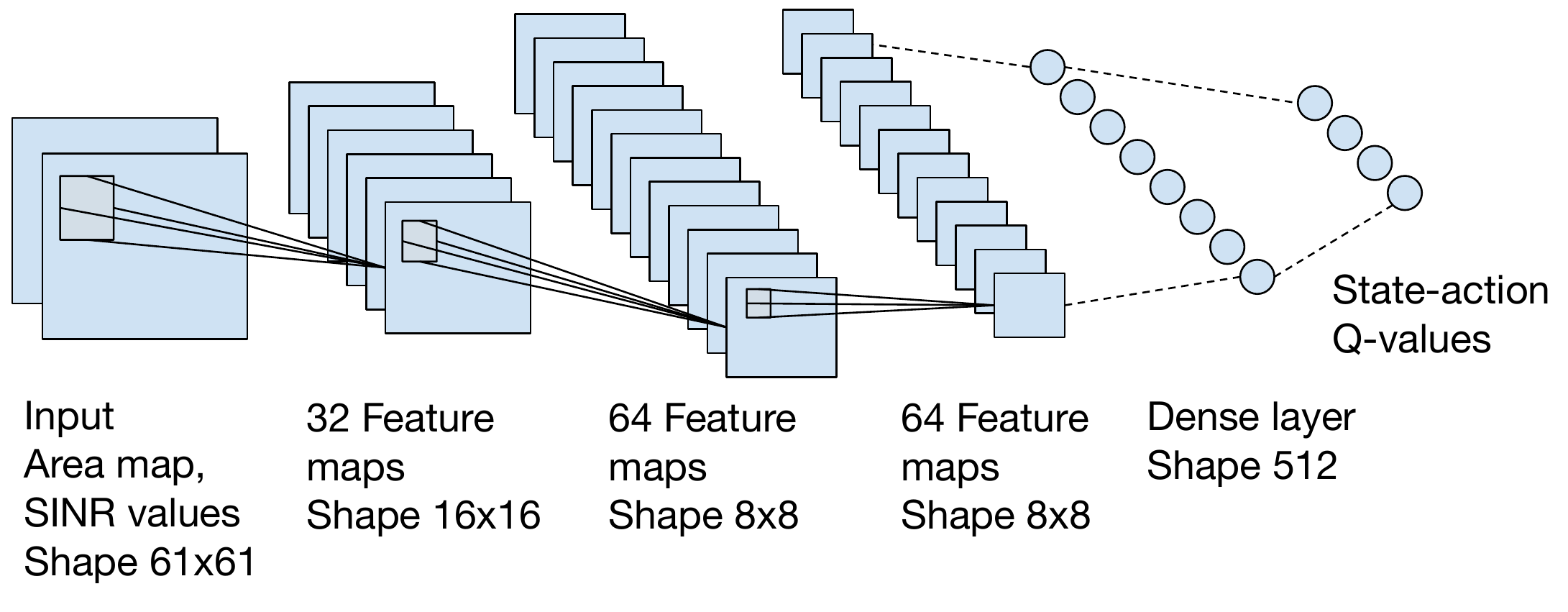}
	\caption{Neural network model used as the Q-network.}
	\label{fig:dqn}
\vspace{-10pt}
\end{figure}
With our problem casted as a PO-MDP, we can apply the deep Q-learning algorithm. For our application, we utilize double Q-learning \cite{van2016deep} and dueling networks \cite{wang2015dueling} extensions to the base DQN algorithm. 
The choice of the neural network model used as the Q-network depends on the application and therefore it needs to be carefully selected for optimal performance. The neural network model we used is shown in Figure \ref{fig:dqn}. 
At the input, there are two 2D arrays, corresponding to the SINR and topology observations described in the previous subsection. 
As displayed, we use 3 convolutional layers with varying number of filters and with each layer having a different filter size. 
%Convolutional layers enable us to process a complex 2D input while still being tractable to train.
There are two fully connected layers, with the final layer output corresponding to the Q-value for each of the possible actions. 
We use ReLU as the activation function after each layer prior to the last layer.
The layer enabling the dueling networks extension is located before the final layer, however we do not show it in the figure for clearer presentation.  
Additionally, we used batch normalization and dropout with probability $p_D$ to accelerate the training of the neural network and for regularization purposes.

\section{Simulation environment}\label{sec:system_model}
\label{sec:dataset}
\begin{figure}[t]
	\centering
	\includegraphics[width=0.40\textwidth]{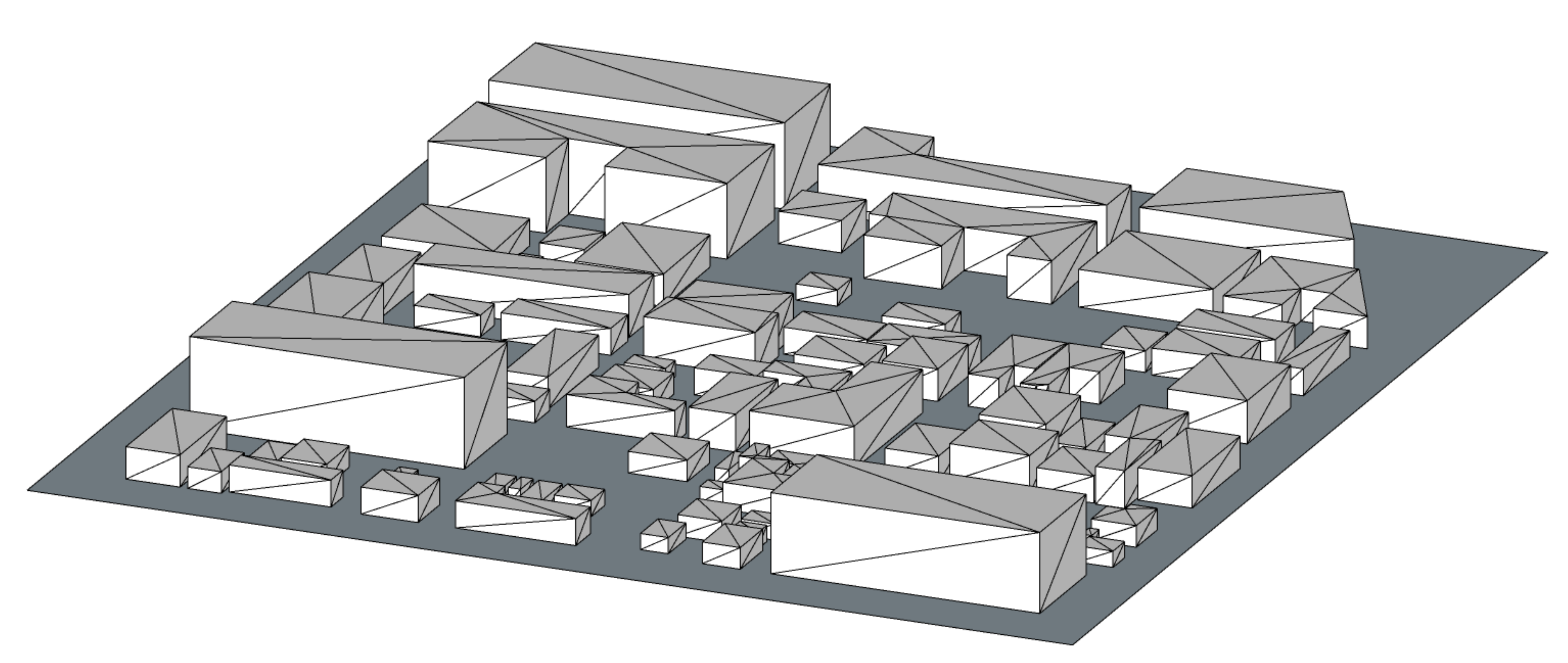}
	\caption{The training urban environment. Approximate size: 550x500 m.}
	\label{fig:train-env}
\end{figure}

\begin{figure}[t]
	\centering
	\includegraphics[width=0.40\textwidth]{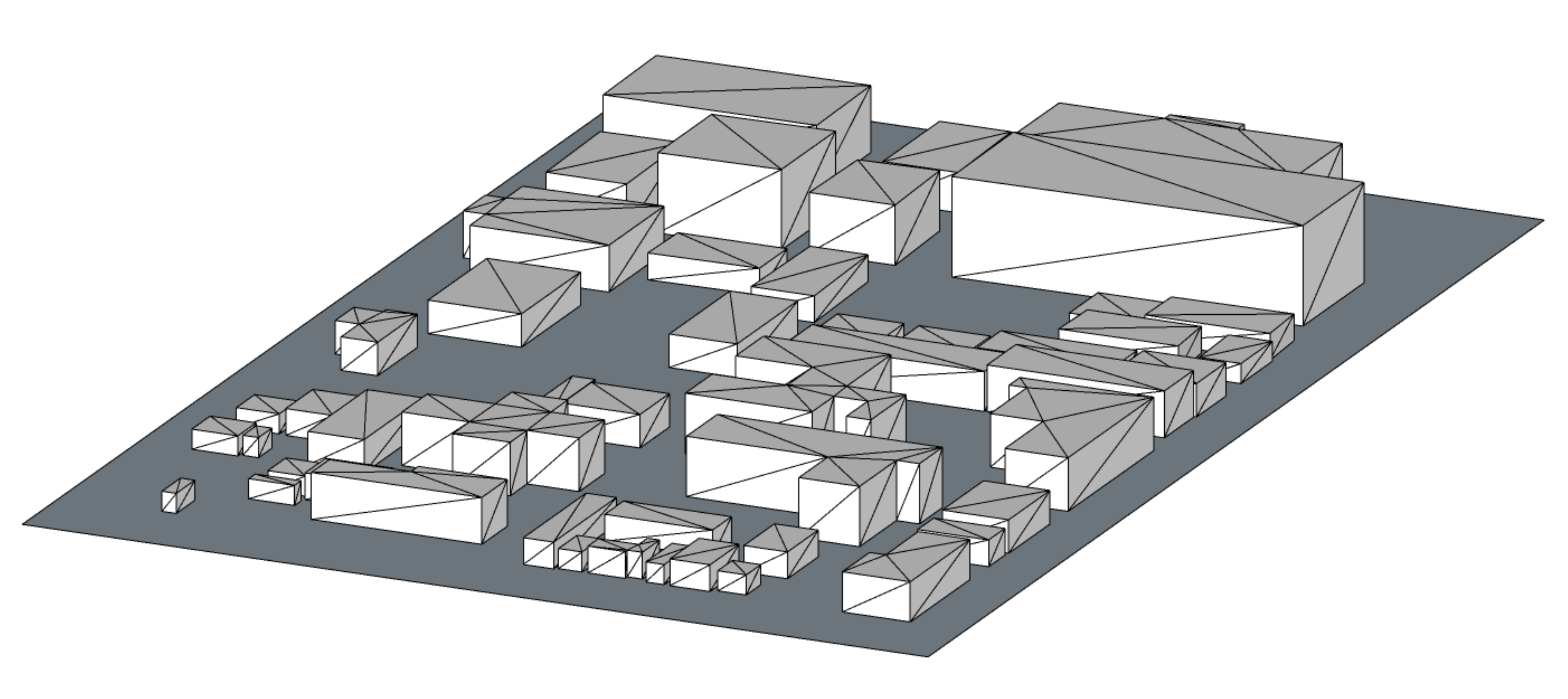}
	\caption{The testing urban environment. Approximate size: 400x500 m.}
	\label{fig:test-env}
\vspace{-10pt}
\end{figure}

We use two separate environments for the training and the testing of our algorithm, shown in Figure \ref{fig:train-env} and Figure \ref{fig:test-env}, respectively. 
Both spaces are meant to resemble a typical medium-elevation urban area.

In order to create realistic conditions to train our Deep RL model, we used a ray-tracing software called Wireless InSite \cite{remcom} to emulate the wireless channel. 
For a given user on the ground we measure the SINR across a uniform grid of points at a fixed height that corresponds to the UAV flying altitude. 
The grid points are separated by 4 meters and they span the entire environment. 
The UAV altitude is set to 10 meters. 
To generate the training data, the user radio was placed at 27 locations uniformly spanning the training environment, while for the testing data we placed the user at 25 different locations in the testing environment. 
The numbers were decided such that user locations uniformly cover the entire space, while still taking a feasible amount of time to make calculations for in the ray tracing software. 
The users transmit a narrow-band signal of 20 dBm power using a frequency of 800 MHz. We introduce a Gaussian noise and a Gaussian background interference across the space with the average combined power of -104 dBm.
Half-wave dipole antennas with vertical orientation are
used at the user and the UAV.

The SINR measurements were then exported and used to build a training and a testing environment in software that the DQN algorithm can interact with. 
At each realization or episode of the environment we use the SINR measurements for a random user location and the UAV is placed at a random location on the grid.
The locations that the UAV can visit correspond to the ones where SINR measurements were recorded. The episode finishes if the UAV reaches the required SINR or the maximum number of steps that the UAV can take is exceeded.

%\begin{figure}[t!]
%	\centering
%	\includegraphics[width=0.35\textwidth]{figures/rays.jpg}
%	\caption{An example heatmap of the SINR values generated for a particular ground user.}
%	\label{fig:heatmap}
%\end{figure}


\section{Results}
In the first part of this section we describe the details of the training of our algorithm and demonstrate that it learns how to move the UAV in order to increase SINR. In the second part, we validate the performance of the algorithm in the testing environment and compare it to a genie algorithm in terms of steps made until convergence to the required SINR.    
\label{sec:results}
\begin{figure}[t]
	\centering
	\includegraphics[width=0.45\textwidth]{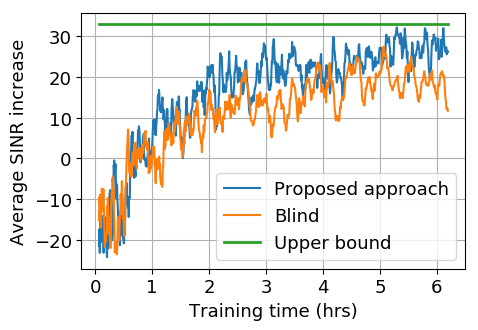}
	\caption{The training results for the proposed model and blind model that does not rely on topology information.}
	\label{fig:train-res}
\vspace{-15pt}
\end{figure}

\subsection{Training}
We train the DQN algorithm in the training environment described in the previous section. 
The maximum number of steps during an episode $t_{\text{MAX}}$ was set to 800 and the target SINR $P_T$, was set to 5 dB. 
When deploying the UAV on the map we ensure that it is not placed in a dead zone with no signal reception, as this would be outside of our problem statement.
In the ray-tracing simulator, these regions occur when there are no direct or reflected paths that can reach the UAV.
For regularization purposes, we rotate the coordinate system of the map by a random multiple of $90^{\circ}$ every training episode.
This ensures that the algorithm does not become biased towards moving in any particular direction over the course of the training.

We use the $\epsilon$-greedy policy for exploration, however the agent's random actions are steered. 
Namely, the agent never takes a random action that would lead to it leaving the map or colliding with a building. Furthermore, the agent repeats the action it has taken in the previous step at probability $0.4\epsilon$ and takes any random allowed action at probability $0.6\epsilon$. 
The repeated movements lead to the agent exploring a larger area through random walk in the early training stage, instead of staying confined to the local space around the starting location. 
The optimal action $\text{argmax}_a~Q_\theta(s,a)$ is taken at probability $1-\epsilon$.
We also ensure that the agent never leaves the map or collides with a building when taking actions according to the DQN. This is done by choosing an action that gives the highest Q-value while still being a legal movement in the environment.
As the UAV moves around, its experience samples are stored in a replay buffer that can store up to $5\times10^5$ samples and when this limit is reached the oldest samples are thrown out to make space for the new ones. 
\begin{table}[b]
\vspace{-14pt}
\renewcommand{\arraystretch}{1.3}
\caption{DQN parameter values used in training}
\label{table:params}
\centering

\begin{tabular}{c||c}
\hline 
Description & Parameter\tabularnewline
\hline 
\hline 
Exploration reward & $c_{E}=1.2$\tabularnewline
\hline 
Discount factor & $\gamma=0.99$\tabularnewline
\hline 
Training batch size & $B_{L}=20$\tabularnewline
\hline 
Training interval & $\tau_{L}=3$\tabularnewline
\hline 
Dropout probability & $p_{D}=0.4$\tabularnewline
\hline 

\end{tabular}

\end{table}

The parameter values used for the training of the DQN algorithm are shown in Table \ref{table:params}. They were selected after tuning for best performance. During minimization we employ gradient clipping and also anneal the learning rate. The width and length of the observation is 61 grid points or 244 m. The movement step size $d_S$ was 4 m.

The training results are shown in Figure \ref{fig:train-res}. To keep track of the progress of training, we measure the average SINR increase from the SINR at the start of the episode to the SINR at the end of the episode, over the most recent 100 episodes. 

To demonstrate the benefits of using 3D maps we evaluate a DQN algorithm that doesn't rely on the 3D map but is otherwise identical to our proposed algorithm. We refer to this algorithm as the `blind' algorithm. The blind algorithm only has a 2D SINR array as an input and the regions blocked by buildings are not populated by $P_L$ but instead left as $P_H$. In training, we use the same parameter settings for the blind and the proposed algorithm. 

Furthermore, we include an upper bound on the mean SINR increase in the training stage. It is calculated by taking the average of the SINR differences between the SINR at all possible starting UAV locations and $P_T$, for all user locations.
This is an upper bound on the mean performance across a large number of episodes.

The results in Fig. \ref{fig:train-res} show that the learning capacity of our proposed algorithm is larger than that of the blind algorithm. 
The intuitive explanation for this is that the algorithm with the knowledge of the topology is more efficient in exploring the space because it can eliminate obstructed areas and because the building knowledge combined with SINR measurements can give it indication where to move to find better SINR.  
The performance curves are noisy due to the nature of training through experience replay and due to the fact that over a 100 episodes the algorithm only experiences a subset of the training environment, which makes the difficulty vary as some user locations are harder to find optimal paths for than others.

\subsection{Testing}
In the testing stage, the algorithm is placed in an entirely new environment and relies only on the trained neural network $Q_\theta$ to guide the movement of the UAV. We use the copy of $Q_\theta$ that had the highest performance during training.
We also introduce a small amount of randomness in decision making, which we found to lead to a better performance. 
The agent takes a random action at probability 0.096. The value of $P_T$ for the testing case was again set to 5 dB and the maximum number of steps $t_{\text{MAX}}$ was reduced to 500, since the testing environment is smaller than the training environment. 

\begin{figure}[t]
	\centering
	\includegraphics[width=0.47\textwidth]{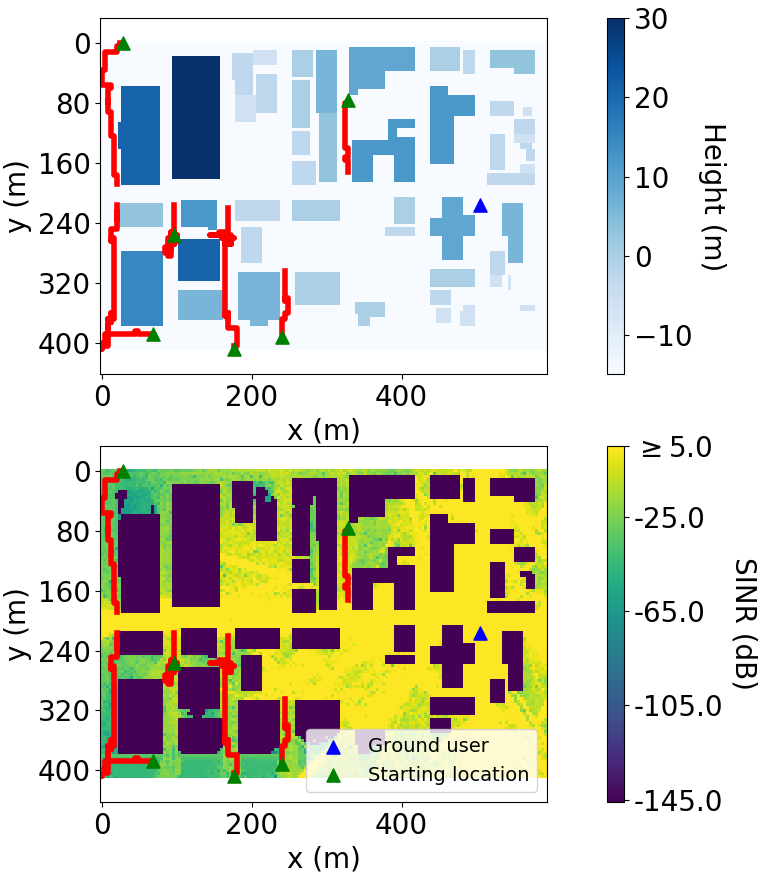}
	\caption{Successful trajectories of the UAV moving according to our algorithm. The upper figure shows the building height relative to the UAV altitude and the lower figure shows the heatmap of the SINR across the map. The green triangle markers represent the starting positions of the UAV. The blue marker represents the location of the user on the ground.}
	\label{fig:renders}

\end{figure}

In Fig. \ref{fig:renders}, we show a number of trajectories of the UAV moving according to our algorithm, where the target SINR was reached in less than $t_{\text{MAX}}$ steps.
The upper figure shows the building height relative to the UAV altitude and the lower figure shows the heatmap of the SINR across the map. 
The user location is the same in each episode, and we place the UAV at random locations.
We can see that the algorithm is capable of following the direction that leads it to improving the SINR and it is also capable of correcting itself when realizing that the current direction of motion is not leading it to better SINR. In the instances where the UAV loops around its location, we can infer that the algorithm explores several path options before settling on the one it deems optimal. Buildings are avoided by the algorithm and it can be inferred that the algorithm eliminates path options because of them. An evidence for the latter is that the UAV tends to move parallel to the building edges, for example. 

\begin{figure}[t]
	\centering
	\includegraphics[width=0.45\textwidth]{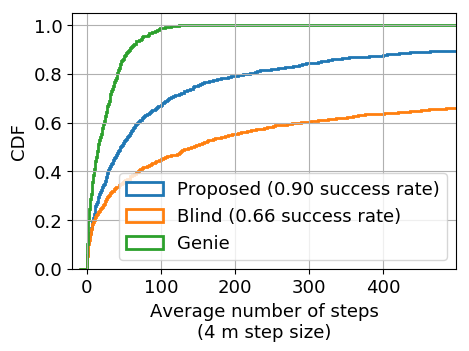}
	\vspace{-3pt}
	\caption{The CDF of the number of steps until convergence to the sufficient SINR for the proposed approach, the blind approach and the genie algorithm.}
	\label{fig:cdf}

\end{figure}

Finally, we analyze the performance of our algorithm over many realizations. 
To get a reference on how fast our algorithm converges to an optimal point we used a genie algorithm for optimal placement. 
The genie algorithm has a complete knowledge of the SINR distribution and building topology, and uses dynamic programming to find the shortest path to a location with sufficient SINR. 
We run the proposed and the blind algorithm for 1000 realizations across the entire testing data set over different user positions.
The results for the number of steps made until convergence to the sufficient SINR for all three algorithms are shown in Fig. \ref{fig:cdf}. The median number of steps until convergence for the proposed algorithm is 44 and 69 for the blind algorithm. 
Furthermore, the proposed algorithm is $90\%$ successful in under $t_{\text{MAX}}$ steps compared to $66\%$ of the blind algorithm. 
Therefore, we show that the knowledge of topology map assists our algorithm even in a novel environment.  
The median number of steps required for the genie algorithm is 14. 
The difference in the number of steps required relative to the proposed algorithm is due to the fact that our algorithm has to explore the space to find good SINR since it does not where the points with sufficient SINR are a priori.  
\section{Conclusions}
In this paper, we used deep reinforcement learning methods to optimize the placement of a UAV communicating to the user on the ground. We consider the case where the ground user location is not known and use topology data to replace statistical models of the channel. We were able to achieve 90\% success rate in moving the UAV to a location that has a sufficient SINR within a limited number of steps. Moreover, our reinforcement learning approach stands out in that it can be applied in any urban environment. Our future work will focus on the scenario where the ground user is moving over the course of optimization. Furthermore, we will explore how the deep reinforcement learning techniques can be used to simultaneously optimize the locations of multiple UAVs serving multiple users on the ground.

\label{sec:conclusion}
%\section{Conclusions}
%\begin{figure}[t!]
%	\centering
%	\includegraphics[width=0.5\textwidth]{figures/performance.png}
%	\caption{Performance of the algorithm after 7 million training steps.}
%	\label{fig:performance}
%\end{figure}
\bibliography{references}
\bibliographystyle{ieeetr}

\end{document}